\documentclass[10pt,a4paper,aip,apl,reprint,nobibnotes,floatfix,superscriptaddress,showpacs]{revtex4-1}

\usepackage[utf8]{inputenc}
\usepackage[english]{babel}

\usepackage{amsfonts}
\usepackage{amsmath}
\usepackage{amssymb}
\usepackage{bbold}
\usepackage[dvipsnames,svgnames,table]{xcolor}
\usepackage{graphicx}
\usepackage{fancyhdr}
\usepackage{float}
\usepackage{braket}
\usepackage{float}
\usepackage[caption=false]{subfig}
\usepackage{tikz}  % drawings
\usepackage{chngcntr}
\usepackage{ulem}

\usepackage[dvipsnames]{xcolor}
\usepackage{hyperref}
\hypersetup{
    pdfstartview={FitH},% fits the width of the page to the window
    colorlinks=true,    % false: boxed links; true: colored links
    linkcolor=NavyBlue, % color of internal links
    citecolor=Maroon,   % color of links to bibliography
    filecolor=NavyBlue, % color of file links
    urlcolor=NavyBlue   % color of external links
}

\makeatletter
\def\footnoterule{\kern -10pt
    \hrule \@width 100pt \kern 10pt} % the \hrule is .4pt high
\makeatother

\begin{document}

%\preprint{AIP/123-QED}

\title[]{Bloch point nanospheres for the design of magnetic traps}
% Force line breaks with \\

\author{F. Tejo}
 \affiliation{Escuela de Ingeniería, Universidad Central de Chile, Avda. Santa Isabel 1186, 8330601, Santiago, Chile }

 \author{C. Zambrano-Rabanal}
 \affiliation{Departamento de Ciencias Físicas, Universidad de La Frontera, Casilla 54-D, Temuco, Chile}

\author{V. L. Carvalho-Santos}
\affiliation{%
Universidade Federal de Vi\c cosa, Departamento de F\'isica, Avenida Peter Henry Rolfs s/n, 36570-000, Vi\c cosa, MG, Brasil. %\\This line break forced% with \\
}%

\author{N. Vidal-Silva}
\email{nicolas.vidal@ufrontera.cl}%
\affiliation{Departamento de Ciencias Físicas, Universidad de La Frontera, Casilla 54-D, Temuco, Chile}

% \homepage{http://www.Second.institution.edu/~Charlie.Author.}

\date{\today}% It is always \today, today,
             %  but any date may be explicitly specified

\begin{abstract}
Through micromagnetic simulations, this work analyzes the stability of Bloch points in magnetic nanospheres and the possibility of using an array of such particles to compose a system with the features of a magnetic trap. We show that a BP can be nucleated as a metastable configuration in a relatively wide range of the nanosphere radius compared to a quasi-uniform and vortex state. We also show that the stabilized Bloch point generates a quadrupolar magnetic field outside it, from which we analyze the field profile of different arrays of these nanospheres to show that the obtained magnetic field shares the features of magnetic traps. Some of the highlights of the proposed magnetic traps rely on the magnetic field gradients achieved, which are orders of magnitude higher than standard magnetic traps, and allow three-dimensional trapping. Our results could be useful in trapping particles through the intrinsic magnetization of ferromagnetic nanoparticles while avoiding the commonly used mechanisms associated with Joule heating.

\end{abstract}

\maketitle

Several propositions for applications of magnetic nanoparticles in spintronic-based devices demand the spin transport electronics of magnetic textures through magnetic fields or electric currents without moving the particle itself \cite{Shinjo-Book,Hirohata-JMMM,Hrcak,Vander-JPD,Goolap-SciRep,Torrejon-Nat,Grolier-Nat,Parkin-Nat,Parkin-Nat2}. Nevertheless, manipulating and moving nanomagnets through external magnetic fields without changing the magnetic pattern of the system also generates exciting possibilities for a plethora of applications \cite{CellMark,Drug1,Drug2,Drug3,Contrast,Hyp1,Hyp2,Hyp3,Hyp4,Kim-Nat}. Within such propositions, an emergent possibility of applying magnetic nanoparticles is using their generated magnetostatic fields as magnetic traps (MTs) \cite{Trap1}, which consists of a system that uses a gradient of the magnetic field to confine charged or neutral particles with magnetic moments \cite{Golub,Kugler,Artetal,Kral,Anderson,Pritchard,Bradley,Gorgier-NJP}, levitate magnetic nanoparticles \cite{Kustura-PRB}, and pinning neutral atoms in low temperatures for quantum storage \cite{FORTAG-RMP,Briegel,Henriet,Singh-Laser}.

MTs generally present a set of devices arranged to generate a quadrupolar magnetic field \cite{Quad1,Quad2,Quad3}. These field profiles can be obtained, for instance, by two ferromagnetic bars parallel to each other, with the north pole of one next to the south of the other. The same field profile can be generated by two spaced coils with currents in opposite directions or four pole tips, with two opposing magnetic north poles and two opposing magnetic south poles \cite{Quad2}. The magnetic field gradient of a quadrupole has the particularity of allowing atoms to leave from the MT due to the zero field strength located at its center \cite{Bergman,Sukumar}. Several solutions to avoid the particles escaping from the trap suggest adding a set of magnetic fields generated by an array of electric currents \cite{Anderson,Pritchard,Trap1,Trap2,Vuletik-PRL,Jian-JPB,Quad3,Roy-SciRep,Luo-NJP,Singh-JAP} to the quadrupolar field. The magnetic fields generated by these electric current distributions ($I$) scale as $I/s$, while their gradient and second derivatives scale as $I/s^2$ and $I/s^3$, respectively \cite{Weinstein}. Here, $s$ represents the characteristic length of the system. In this context, the smaller these MTs, the better the particle confinement, and several techniques to diminish their sizes were developed \cite{Weinstein,Fortagh,Hess,Raab,Folman,Muller,Dekker,Hansel,Ott}. However, the miniaturization of MTs using an array of nanowires and coils for manipulating atoms faces the problem of energy dissipation by Joule heating \cite{Potting}. In this context, the intrinsic dipolar fields of specific magnetic textures of ferromagnetic nanoparticles emerge as natural candidates to compose nanosized MTs \cite{Amir-PS,West2012,Allwood-2006}.

A promising proposition to adopt nanosized magnetic textures as sources of magnetic field gradient is using the magnetostatic field generated by spin textures in chiral magnets \cite{Skyrmion1,Skyrmion2,Skyrmion3}. Indeed, because the magnetostatic field generated by a skyrmion lattice is similar to that created by two helices carried by electric currents \cite{Skyrmion1}, nanoscaled MTs can be engineered by stacking chiral ferromagnets hosting skyrmions \cite{Skyrmion2,Skyrmion3}. Another exciting result regarding magnetostatic fields produced by topological spin textures is the generation of a quadrupolar field by just one magnetic nanoelement, as evidenced by Zambrano \textit{et. al.} \cite{Zambrano-SciRep} for a magnetic nanosphere hosting a Bloch point (BP). Nevertheless, in that case, the nanosphere is located at the center of the quadrupolar field, reducing the feasibility of applying this only structure as a magnetic trap. Following these ideas and motivated by the proposition of stacking skyrmion lattices to compose MTs, we analyze, through micromagnetic simulations, the possibility of using a BP array as an MT. We start by exploring the stability of a BP on a nanosphere as a function of their geometrical and magnetic parameters. After determining the magnetostatic field of a BP, we show that an array of four BP nanospheres generate a magnetic field gradient with all properties to be applied as an MT. 

Our main focus is presenting a proposition to use BP nanospheres as sources of magnetic fields in MTs. Therefore, we obtain the stable and metastable states of a ferromagnetic nanosphere as a function of its radius, $R$, and magnetic parameters. The analysis is performed through micromagnetic simulations using the OOMMF code \cite{oommf}, a well-known tool that agrees well with experimental results on describing the magnetization of nanoparticles. In the simulations, we consider three values to $M_s$ and the exchange stiffness, $A$, characterizing iron ($M_s \approx 1700$ kA/m and $A = 21$ pJ/m), Permalloy ($Ms \approx 850$ kA/m and $A = 13$ pJ/m), and cobalt ($Ms \approx 1450$ kA/m and $A = 56$ pJ/m). To simulate a smooth spherical geometry, we consider a cubic cell with the size of $0.5\times0.5\times0.5$ nm$^3$. 

The local and global minima are obtained by comparing the total energy, $E$, of three magnetic profiles: quasi-uniform, where the magnetic moments slightly deviate from the purely parallel direction \cite{Landeros,Vagson}; vortex, characterized by a curling magnetization field around an out-of-plane core \cite{Riveros2016}; and  BP configuration, characterized by two magnetic bobbers \cite{bobbers} separated by a texture that, in a closed surface around its center, the magnetization field covers the solid angle an integer number of times \cite{MaloSlo}. These magnetic patterns are obtained by relaxing the system from three different configurations and determining the total energy, $E=E_x+E_d$, of the relaxed state. Here, $E_x$ and $E_d$ are the exchange and dipolar contributions to the total energy. The first initial state consists of a single domain, which, after relaxation, reaches a quasi-uniform configuration. The second and third initial configurations consist of a rigid vortex and BP artificially imposed. Subsequently, both states let it relax to achieve a vortex and a BP as metastable system configurations, respectively. 

\begin{figure}[ht]
\includegraphics[width=8cm]{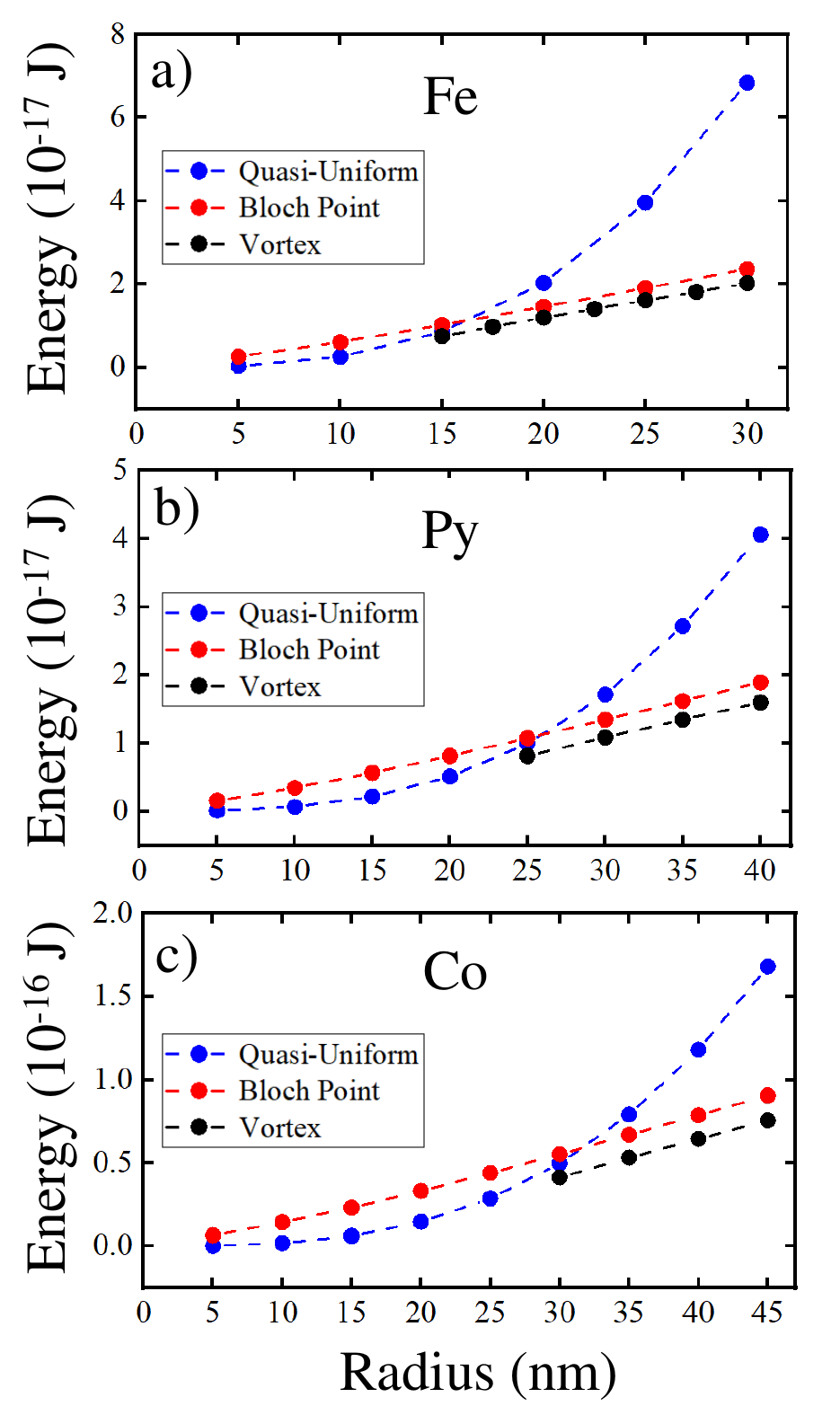}
\caption{Magnetic energy of the quasi-uniform (blue dots), BP (red dots), and vortex (black dots) configurations as a function of the nanosphere radius and different materials.}
\label{fig:1}
\end{figure}

The energies of final states for a nanosphere of Fe, Py, and Co are shown in  Fig. \ref{fig:1}. One notices that due to the role that the exchange interaction plays in systems with small sizes, the quasi-uniform state appears as groundstate when the nanosphere radius is smaller than a threshold value of $R_c\approx15$ nm (Fe), $R_c\approx25$ nm (Py), and $R_c\approx30$ nm (Co). Nevertheless, the contribution of dipolar energy increases with the system size, and at these threshold values, both the BP and the vortex become energetically favorable. Indeed, one can notice that the vortex configuration corresponds to the groundstate, while the BP has a slightly higher energy. As a result, the BP configuration is then a metastable state, whereas the vortex is the more stable state. Therefore, we claim that under certain conditions, a BP can be stabilized and conclude that in addition to its topological protection, the BP also has energetic metastability, compared to a quasi-uniform state, for radii greater than the material-dependent threshold value. To diminish computational effort, we will focus our discussion on a Fe nanosphere with $R=15$ nm, which is the lower limit to the critical radius allowing the BP metastability, and it is appreciated possesses the minimum energy difference with the vortex configuration. Nevertheless, no qualitative changes for the results presented here should be observed if we consider Py or Co nanospheres hosting BPs. 

\begin{figure}[ht]
\includegraphics[width=8.8cm]{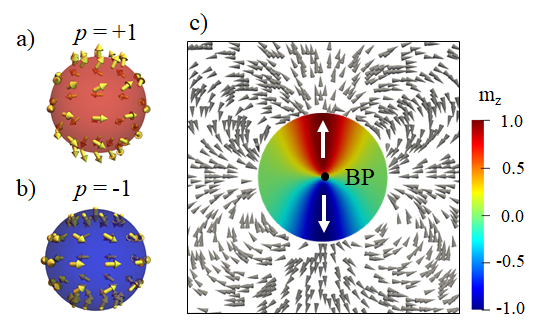}
\caption{Bloch point profiles obtained from Eq. \eqref{Ansatz} for $\gamma=\pi/2$ and (a) $p=+1$ and (b) $p=-1$. (c) Representation of the quadrupolar magnetostatic field generated from a BP nanosphere projected in a plane and obtained from micromagnetic simulations. The color bar shows the magnetization component pointing in the $z$-axis direction.}
\label{fig:2}
\end{figure}

After showing that BPs can appear as metastable states compared to quasi-uniform and vortex configurations, we analyze the properties of the magnetostatic field of such a system. The vector field of a BP can be parameterized by the normalized magnetization written in spherical polar coordinates as ${\bf M}/M_s =\left(\sin\Theta\cos\Phi,\sin\Theta\sin\Phi,\cos\Theta\right)$, where $M_s$ is the magnetization saturation. Under this framework, the magnetic profile of a BP configuration can be modeled with the ansatz \cite{Elias-EPL} 

\begin{equation}\label{Ansatz}
    \Theta(\theta) = p\theta+\pi(1-p)/2\,\hspace{0.5cm} \text{and}\hspace{0.5cm} \Phi(\phi) = \phi+\gamma.
\end{equation}
    
Here, $\theta$ and $\phi$ are the standard polar and azimuthal angles describing the spherical coordinates, and $p=\pm 1$ is the BP polarity, which determines the orientation of the magnetic moments in nanosphere poles in the $z$-axis direction. In this case, the magnetic moments point outward or inward for $p=+1$ and $p=-1$, respectively, as depicted in Figs. \ref{fig:2}a) and b). The parameter $\gamma$ accounts for determining the BP helicity. For instance, $\gamma=0$ represents a hedgehog magnetization field pointing outward the sphere center, while $\gamma=\pi/2$ depicts a tangent-to-surface configuration in the sphere equator. The ansatz (\ref{Ansatz}) has been previously used to determine the magnetostatic field outside a BP nanosphere \cite{Zambrano-SciRep}, given by

\begin{equation}
{\bf H}(r,\theta)=\frac{M_sR^4}{48r^4}(1-\cos\gamma)
\left[2\,P_2(\cos\theta)\,{\hat r}
    +\sin 2\theta\,{\hat\theta}
\right]\,,
\label{Hmagout}
\end{equation}

\noindent
where $P_2(x)$ is the Legendre polynomial of degree 2, and $r$ is the radial component of the position of a point outside the nanosphere. From the BP nanosphere property that $\gamma$ adopts a constant quasi-tangential configuration in the nanosphere equator \cite{Elias-EPL,Pyly-PRB,Tejo-SciRep}, one observes that the magnetostatic field outside the considered system consists of a quadrupole, which is consistent with Eq. \eqref{Hmagout}, and is also obtained in our micromagnetic simulations, as shown in Fig. \ref{fig:2}c). 

\begin{figure*}[ht]
\includegraphics[width=18cm]{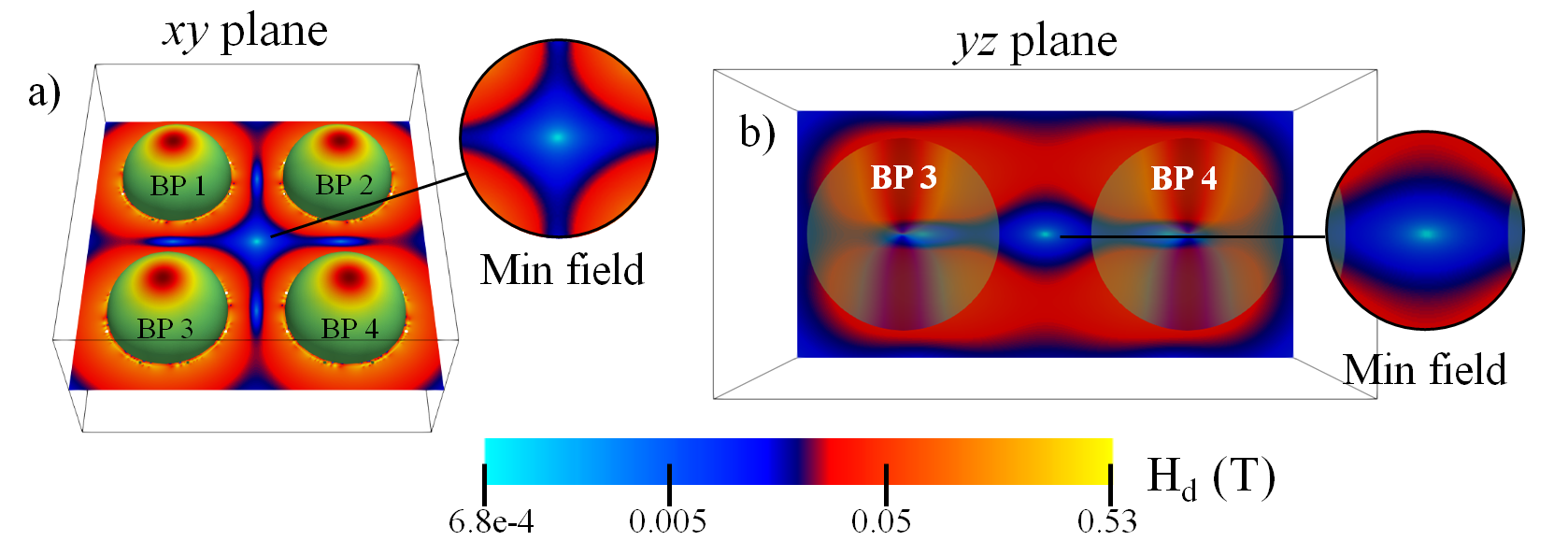}
\caption{Magnitude of the Magnetostatic field ($H_d$) profile of the proposed distribution as potential magnetic traps corresponding to Array I. a) depicts the field distribution along the $xy$ plane, and b) along the $yz$ plane. The color bar represents the strength of $H_d$.}
\label{fig:3}
\end{figure*}
Although this field profile seems to be a good candidate for MTs, the nanosphere is located at the center of the quadrupolar field, which avoids using this only structure for this application. Therefore, we discuss on the possibility of using an array of such elements to generate a magnetic field gradient with the features of an MT. The proposed arrays consist of four Fe BP nanospheres with a radius of 15 nm. These nanospheres are symmetrically positioned in the vertices of a square inside a rectangular prism with dimensions $120\times120\times60$ m$^3$ (see Fig. \ref{fig:3}-a)). The proposed arrays differ by the square side size and the BP polarities as presented in table \ref{T1}, where $p_i$ refers to the BP polarity in the vertex $i$. It is important to point out that in all the simulated arrays, the chirality acquired by the BPs emerges as a consequence of the energy minimization \cite{Zambrano-SciRep}.

\begin{table}[htp]
\caption{Analyzed arrays}
\begin{center}
\begin{tabular}{|c|c|c|}
\hline
Array & Size side & BP polarities \\
 \hline
 I & 60 nm & $p_1=p_2=p_3=p_4=1$ \\
 II & 60 nm & $p_1=p_2=-1$ and $p_3=p_4=1$ \\
 III & 60 nm &$p_1=p_4=-1$ and $p_2=p_3=1$ \\
 \hline
\end{tabular}
\end{center}
\label{T1}
\end{table}%

\begin{figure}[ht]
\includegraphics[width=8.5 cm]{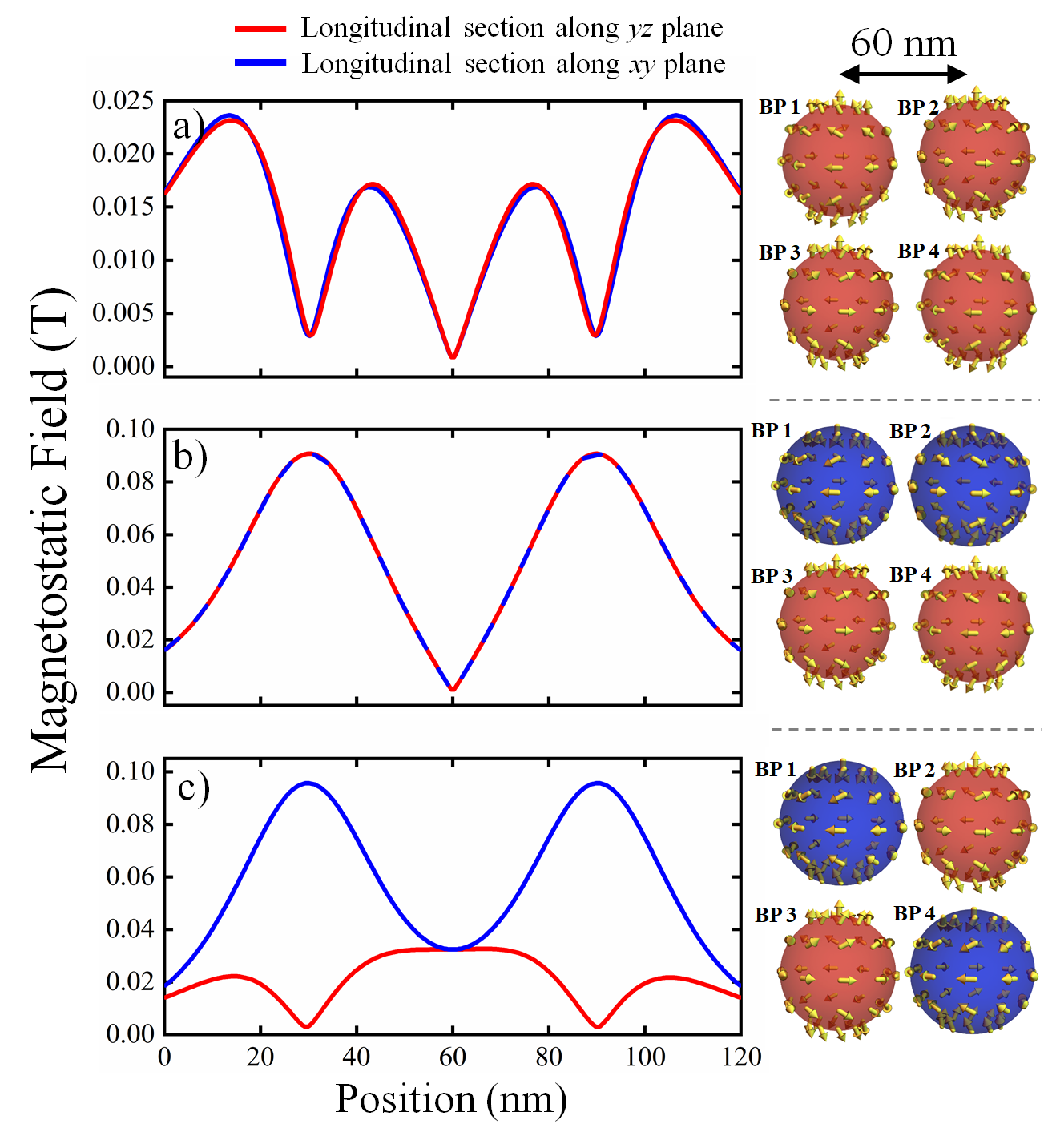}
\caption{Field strength of a) Array I, b) Array II, and c) Array III in the longitudinal sections crossing the center of $xy$ and $yz$ planes.}
\label{fig:4}
\end{figure}

%\begin{figure}[ht]
%\includegraphics[width=8.5cm]{ArrayI.png}
%\caption{Field strength of a) Array I, b) Array III, and c) Array IV in the longitudinal sections crossing the center of $xy$ and $yz$ planes.}
%\label{fig:4}
%\end{figure}

Firstly, we analyze the profile of the magnetic field of Array I. Main results are summarized in Fig. \ref{fig:3} and Fig. \ref{fig:4}a). In the former, we present the snapshots of the modulus of the magnetostatic field ($H_d$) profile in the $xy$ and $yz$ planes, respectively, in a longitudinal section of these planes. The color map of the magnetic field allows us to notice that Array I gives place to a range of magnetic fields going from $H_d \approx 0.5$ T, in the regions surrounding the nanospheres, until a minimum value of $H_d \approx 6.8 \times 10^{-4}$ T in the center of the array, as shown in Figs. \ref{fig:3}a) and b). The detailed analysis presented in Fig. \ref{fig:4}a) of the field profile in the longitudinal sections in the $xy$ and $yz$ planes reveals that the magnetic field generated by Array I presents local and global minima depending on the position. While the local minima occur in the center of two adjacent nanospheres, the global one is in the system center. Therefore, the presented results show the existence of a magnetic field gradient in space, which can be numerically determined. We obtain that the field gradients are in the order of $\sim 10^5-10^6$ T/m, much higher than the field gradient of conventional MTs \cite{Weinstein,Reichel}. The existence of high gradients of magnetic fields yields narrower confinement, making systems with this property very interesting for applications in MTs \cite{Weinstein}. Also, the similar behavior of the magnetic field in both longitudinal sections allows the symmetric confinement in three different places (local and global minima of magnetic fields) of particles if they are charged from $x$ or $y$ axes. Finally, local minima have the advantage of ensuring higher stability to the trapped particles. 

%We highlight that Array II presents the same qualitative behavior as Array I. The Main differences are the localization of local minima and the field gradients, which are in the $\sim 10^4-10^5$ T/m range. Therefore, MTs based on Array II would have smaller stability in confining particles.

%\begin{figure}[ht]
%\includegraphics[width=8.5cm]{ArrayIII.png}
%\caption{Field strength of Array III in the longitudinal sections crossing the center of $xy$ and $yz$ planes.}
%\label{fig:5}
%\end{figure}

Because changing the distance among the nanoparticles affects the strength of the magnetostatic field \cite{Zambrano-SciRep}, we also propose changes in the structure of the array. Therefore, we analyze the field profile when the nanosphere polarity distribution is given by Array II. Fig. \ref{fig:4}b) shows the field distribution and its strength as a function of the position along the longitudinal sections along $xy$ and $yz$ planes. One notices that the generated magnetostatic field has exactly the same behavior in both longitudinal sections, reaching the maximum values in the space between two neighbor spheres ($\approx 30$ nm y $\approx 90$ nm) and a unique global minimum in the array center. The appearance of just one minimum weakens the implementation of Array II as an MT. 
%However, the produced magnetic field gradients are bigger than Array II, so a MT using Array III would present a lower particle loss due to the narrow confinement.

%\begin{figure}[ht]
%\includegraphics[width=8.5cm]{ArrayIV.png}
%\caption{Field strength of Array IV in the longitudinal sections crossing the center of $xy$ and $yz$ planes.}
%\label{fig:6}
%\end{figure}

Finally, we consider the magnetic field generated by Array III, whose results are given in Fig. \ref{fig:4}c). In this case, we obtain that the field profiles of the longitudinal sections along $xy$ and $yz$ planes are different. Indeed, the magnetic field along the $xy$ plane has two maxima between the BP nanospheres 1 and 3, and 2 and 4, and a nonzero minimum in the array center. On the other hand, the field profile along the $yz$ plane presents a maximum value in the array center and two minima between BP nanospheres 1 and 2, and 3 and 4. Therefore, Array III generates a magnetic field with a triple saddle point, and this array does not work as a potential MT since the magnetic field does not have the features to stabilize atoms or particles with magnetic moments. 

The above-described results show that different distributions of magnetic fields are obtained depending on the BP nanosphere polarity distribution. Two of these fields present the features to be used as MTs. The main advantages of using an array of BP nanospheres to generate a gradient of magnetic fields are the lower cost of production when compared to lithographic processes that  use materials such as Al$_2$O$_3$, AIN, Si, and GaAs to fabricate conductor nanowires in a chip \cite{Aldrich}. In addition, the proposed setting also has the advantage of avoiding energy losses due to the heating of the nanospheres. We highlight that although the BPs are metastable states, the increase in the temperature of the MT due to the motion of the trapped particles is not big enough to denucleate the BP from the nanospheres. 

In summary, we have analyzed the magnetostatic properties of magnetic nanospheres hosting a BP as a metastable state. In addition to their topological protection, BPs have energetic metastability in nanospheres with a radius above a threshold value that depends on the material parameters. After discussing the energy of BP nanospheres, we determine the magnetostatic field generated outside it. The micromagnetic simulations reveal the appearance of a quadrupolar field, as previously reported from analytical calculations \cite{Zambrano-SciRep}. We then analyzed the magnetic field profile of different arrays of BP nanospheres to propose the production of a magnetic trap. We showed that the array with the better features to be used as magnetic traps consists of four nanospheres hosting BPs with positive polarities. Although we analyzed the proposal by projecting the magnetostatic field profiles into a given plane, they are essentially three-dimensional quadrupolar fields. This feature adds a new degree of freedom to potential MTs by allowing charging particles from different directions of 3D space.\\

\textit{Acknowledgments}: The work of F.T. was supported by ANID + Fondecyt de Postdoctorado, convocatoria 2022 + Folio 3220527. V.L.C.-S. acknowledges the support of the INCT of Spintronics and Advanced Magnetic Nanostructures (INCT-SpinNanoMag), CNPq 406836/2022-1. V.L.C.-S. also thanks the Brazilian agencies CNPq (Grant No. 305256/2022-0) and Fapemig (Grant No. APQ-00648-22) for financial support. N. V-S acknowledges funding from ANID Fondecyt Iniciacion No. 11220046.\\

Data availability: The data that support the findings of this study are available from the corresponding author upon reasonable request.

\end{document}